\documentclass[aps,pra,twocolumn]{revtex4}
\usepackage[usenames,dvipsnames]{color}

\usepackage[latin1]{inputenc}
\usepackage{graphicx}
\usepackage{amssymb}
\usepackage{subfigure}
\usepackage{stackrel}
\usepackage{amsmath}
\usepackage{amsfonts}
\usepackage{bm}
\usepackage{color}
\newcommand{\ket}[1]{\ensuremath{\left|{#1}\right\rangle}}
\newcommand{\bra}[1]{\ensuremath{\left\langle{#1}\right |}}

\newcommand{\beq}{\begin{equation}}
\newcommand{\eeq}{\end{equation}}
\newcommand{\bse}{\begin{subequations}}
\newcommand{\ese}{\end{subequations}}
\newcommand{\bea}{\begin{eqnarray}}
\newcommand{\eea}{\end{eqnarray}}
\newcommand{\bit}{\begin{itemize}}
\newcommand{\eit}{\end{itemize}}
\newcommand{\bpmatrix}{\begin{pmatrix}}
\newcommand{\epmatrix}{\end{pmatrix}}

\newcommand{\be}{\begin{equation}}
\newcommand{\ee}{\end{equation}}
\newcommand{\ben}{\begin{eqnarray}}
\newcommand{\een}{\end{eqnarray}}

\begin{document}


\title{Dynamics of entanglement in systems of identical fermions undergoing decoherence}


\author{A. Vald\'es-Hern\'andez$^1$, A. P. Majtey$^1$, A. R. Plastino$^2$}
\affiliation{$^1$Instituto de F\'{\i}sica, Universidade Federal do Rio de Janeiro, Caixa
Postal 68528, Rio de Janeiro, RJ 21941-972, Brazil\\$^2$CeBio  y Secretar\'{i}a de Investigaciones, Universidad Nacional del Noroeste de la Prov. de Buenos Aires, UNNOBA-Conicet, Roque Saenz-Peña 456, Junin, Argentina\\}

\begin{abstract}
Information that is stored in quantum-mechanical systems can be easily lost because of the interaction with the environment in a process known as decoherence. Possible physical implementations of many processes in quantum information theory involve systems of identical particles, whence comprehension of the dynamics of entanglement induced by decoherence processes in identical-particle open systems becomes relevant. Here we study the effects, and concomitant entanglement evolution, arising from the interaction between a system of two identical fermions and the environment, for two paradigmatic quantum channels. New entanglement measures are introduced to quantify the entanglement between the different parties, and a study of the dynamics of entanglement for some particular examples is carried out. Our analysis, which includes also the evolution of an entanglement indicator based on an entropic criteria, offers new insights into the dynamics of entanglement in open systems of identical particles, involving the emergence of multipartite genuine entanglement. The results improve our understanding of the phenomenon of decoherence and will provide new strategies to control it.

\end{abstract}


\maketitle

\section{Introduction}
Decoherence is a quantum phenomenon that plays an important role
in connection both with the foundations of quantum physics and
with its technological applications \cite{Luiz,S05,NC00}.
Decoherence is also intimately connected to another key ingredient
of the quantum world: quantum entanglement \cite{BZ06}.
Indeed, the various effects associated with
the phenomenon of decoherence are due to the emergence of
entanglement between the system under study and its
environment. The interaction between an imperfectly isolated system and its surroundings,
leads to the gradual disappearance of several quantum features
exhibited by the system. These effects are at the core of
the nowadays orthodox, decoherence-based explanation of the
emergence of the classical world from quantum physics \cite{JZKGKS03,Zurek03,S05}

Possible physical implementations of most of the processes in quantum information involve systems composed of identical particles \cite{ESBL02}. Just to mention a few, systems such as semiconductor quantum dots, in which charge carriers are confined in all three dimensions \cite{SLM01}, or neutral atoms in magnetic \cite{PRA95} or optical \cite{BBDE01} microtraps. However, contrary to what occurs in multipartite systems of  distinguishable particles --in which much attention has been paid to analyse the dynamics of entanglement (mainly in qubit systems)--, the evolution of entanglement and decoherence in systems composed of indistinguishable particles remains a largely unexplored field. On one hand, the concept of entanglement in these kind of systems exhibits some differences from the corresponding concept as applied to systems consisting of distinguishable subsystems, being perhaps more controversial. However, there is general consensus that in systems of identical fermions the minimum quantum correlations between the particles that are required by the indistinguishability  and the anti-symmetry of the fermionic state do not contribute to the state's entanglement \cite{ESBL02,GMW02,GM04,NV07,BMPSD2012,LV08,PMD09,BPCP08,AFOV08,LNP05,ZP10,ZPCP12,SCKLL01,GM05,oliveira}. On the other hand, a resource theory based on local operations and classical communication (LOCCs) is no longer suitable: Due to the necessary (anti)symmetrization of identical particles' states, the notion of local operations does not apply anymore. 

The aim of the present contribution is thus to advance in the investigation of the dynamics of entanglement and decoherence in an open system composed of identical particles. Specifically, we analyse the effects, and concomitant entanglement evolution, arising from the interaction between a pair of identical fermions and the environment. 

The article is organised as follows. Exchange-symmetry preserving transformations are discussed in section II, where it is evinced that the notion of local operations is foreign to identical-particle systems. Section III contains the preliminaries for the subsequent analysis of the entanglement distribution. First, we present a brief review of the definition and quantification of the entanglement between two identical fermions. Then we define appropriate measures for the entanglement between one fermion and the rest of the system (fermion plus environment), and also between one of the fermions and the environment. In section IV we investigate the dynamics of the entanglement in an open system consisting of two identical fermions. We do this by considering two decoherence processes that are paradigmatic in quantum information theory. Our results open the possibility to study the efficiency of some previously developed entropic entanglement criteria, a task that is carried out in section V. Finally, some conclusions are drawn in section VI. 

\section{Exchange-symmetry preserving transformations}\label{sim}

Consider a bipartite system $S$ composed of a pair of subsystems $a$ and $b$, immersed in an environment $E$. Initially $S$ is in an arbitrary state $\rho_{ab}(0)$, whereas $E$ is assumed to be in a pure state denoted as $\ket{0}_{E}$. The initial density matrix of the
complete system is thus given by 
\begin{equation}
\rho(0)=\rho_{ab}(0)\otimes\ket{0}\bra{0}_E.  \label{rho00}
\end{equation}%
At $t=0$, $S$ starts to interact with the environment $E$ by
means of a unitary transformation $U=\exp{(-iHt/\hbar)}$,  with $H$ the total ($S$ plus $E$) Hamiltonian. As a result, the effective evolution of the bipartite system $ab$ writes as 
\begin{equation}
\rho _{ab}(t)=\textrm{Tr}_E({U\rho(0)U^{\dag}})
=\sum_{\mu }K_{\mu }(t)\rho _{ab}(0)K_{\mu }^{\dag }(t),
\label{kraus}
\end{equation}%
where the $K_{\mu }=\bra{\mu} U\left\vert {0}%
\right\rangle _{E}$ are the Kraus operators associated to the transformation 
$U$, and $\{\ket{\mu}_{E}\}$ is an orthonormal basis of the Hilbert space $\mathcal{H}_{E}$. Since Tr$\rho _{ab}(t)=1$, the Kraus operators satisfy the relation $\sum_{\mu }K_{\mu }K_{\mu }^{\dag
}=\mathbb{I} $. 

We now assume that both subsystems $a$ and $b$ possess the same dimension, say $n$. The $n^2$-dimensional Hilbert space $\mathcal{H}_{S}$ can thus be decomposed into two subspaces, namely $\mathcal{H}_{-}$ with dim$\mathcal{H}_{-}=n(n-1)/2$, and $\mathcal{H}_{+}$ with dim$\mathcal{H}_{+}=n(n+1)/2$, that are spanned by basis vectors that are antisymmetric and symmetric, respectively, under the exchange of the subsystems $a$ and $b$. Let $\{\ket{\psi^{\pm}_{k}}_{S}\}$ be an orthonormal basis of the subspace $\mathcal{H}_{\pm}$, where $k$=$1,2,...n(n\pm1)/2$. If the initial state of the composite system $S$ has a well-defined symmetry under the exchange $a\leftrightarrow b$, it decomposes as \begin{equation}
\rho_{ab}(0)=\sum_{k}p_{k}\ket{\psi^{\pm}_{k}}\bra{\psi^{\pm}_{k}}_{ab}.
\label{rho0}
\end{equation}
Clearly, a necessary and sufficient condition for the evolved state ($\ref{kraus}$) to preserve the initial symmetry is that 
\begin{equation}
[H,P_{ab}\otimes\mathbb{I}_{E}]=0,
\label{conmutador}
\end{equation}
 with $P_{ab}$ the subsystem-exchange operator. 
 
Up to now we have referred to $E$ as the environment in which the bipartite system $ab$ is immersed, without making further assumptions about its nature. In particular, $E$ may also be a bipartite system, composed of independent environments $A$ and $B$. Such decomposition of $E$ allows for a \textit{local evolution}, in which each subsystem $a$ and $b$ can couple independently (or locally) with its own environment $A$ and $B$, respectively. A local evolution corresponds thus to a Hamiltonian of the form
\begin{equation}
H=H_{aA}+H^{\prime}_{bB}
\label{Hlocal}
\end{equation}
where $
H_{aA}=H^{free}_{a}+H^{free}_{A}+H^{int}_{aA},$ and similarly for $H^{\prime}_{bB}$. In this case the unitary evolution factorizes as $U$=$U_{aA}U^{\prime}_{bB}$, and the Kraus operators decompose as $K_{\mu=(\mu_{A},\mu_{B}) }=\bra{\mu_{A}} U_{aA}\ket{0_{A}}\otimes\bra{\mu_{B}} U_{bB}\ket{0_{B}}$. If, on the contrary, $E$ does not decompose into two independent environments but stands for the \textit{common} environment of both (noninteracting) subsystems $a$ and $b$, the total Hamiltonian writes as
\begin{equation}
H=H^{free}_{a}+H^{\prime free}_{b}+H^{free}_{E}+H^{int}_{aE}+H^{\prime int}_{bE}.
\label{Hnonlocal}
\end{equation}
This case corresponds to a \textit{global evolution}, in which $U$ cannot in general be decomposed as $U_{aE}U^{\prime}_{bE}$, and the subsystems $a$ and $b$ do not evolve independently.

Now, it is easy to see that the condition (\ref{conmutador}) holds iff the evolution is global and symmetric, i.e., if $H$ has the estructure (\ref{Hnonlocal}) with $H$=$H^{\prime}$ for all the free and interaction Hamiltonians. Indeed, for 
\begin{equation}
H=H^{free}_{a}+H^{free}_{b}+H^{free}_{E}+H^{int}_{aE}+H^{int}_{bE},
\end{equation}
it is immediate to verify that $(P_{ab}\otimes\mathbb{I}_{E})H(P_{ab}\otimes\mathbb{I}_{E})$=$H$. The converse, that (\ref{conmutador}) implies a global, symmetric evolution, can be verified assuming that the evolution is local and symmetric and showing that the corresponding Hamiltonian, namely
\begin{equation}
H=H^{free}_{a}+H^{free}_{b}+H^{free}_{A}+H^{free}_{B}+H^{int}_{aA}+H^{int}_{bB},
\label{Hloc2}
\end{equation}
does not comply with Eq. (\ref{conmutador}). That this is so follows immediately from the interaction terms $H^{int}_{aA}+H^{int}_{bB}$, which under $P_{ab}\otimes\mathbb{I}_{E}$ transform into $H^{int}_{bA}+H^{int}_{aB}$, thus preventing the invariance of $H$ under the exchange $a\leftrightarrow b$. 

These results show that an open bipartite system preserves the symmetry under the exchange of its (noninteracting) parts $a$ and $b$ if and only if there is a common environment so that the evolution is global (nonlocal). This is specially relevant when studying decoherence processes in identical particle systems, a matter that will be analysed below, in relation with systems of two identical fermions. In fact, in the particular case when $a$ and $b$ are indistinguishable subsystems, clearly the Hamiltonian (\ref{Hlocal}) can be ruled out from the start, since there is no possibility of distinguishing them through an interaction.


\section{Preliminaries}

\subsection{Entanglement in systems of identical fermions}\label{IIIA}

Consider that $a$ and $b$ represent two identical fermions. (Though indistinguishable, in what follows we will often use the notation $a$ and $b$ to refer to  ``one" and ``the other" fermion. Such notation is introduced for clarity purposes, and must not be understood as a labeling that distinguishes between the two fermions). Let $\ket{\phi_{\kappa}}$ and $\ket{\phi_{\kappa^\prime}}$ be two single-fermion states. The antisymmetric combination
\beq\label{slater}
\ket{\psi^{sl}_k}=\frac{1}{\sqrt{2}}(\ket{\phi_\kappa}_{a}\ket{\phi_{\kappa^\prime}}_b-\ket{\phi_{\kappa^\prime}}_{a}\ket{\phi_\kappa}_b).
\eeq
defines what is called a Slater determinant (and is said to have Slater rank 1). A composite system consisting of two identical fermions
is regarded as separable (i.e., non-entangled) if and only
if its density matrix is of the form \cite{GMW02}
\begin{equation}
\rho^{sep}_{ab}=\sum_{k}p_{k}\ket{\psi^{sl}_k}\bra{\psi^{sl}_k}
\label{rhosep},
\end{equation} 
with $\sum_{k}p_{k}=1.$ That is, a pure state of two identical fermions is simply a single Slater determinant, whereas mixed separable states are those that can be expressed as a statistical mixture of pure states of Slater rank 1. Here, by ``entanglement'' we mean entanglement between particles (as opposed to entanglement
between modes). Comparison of Eq. (\ref{rho0}) with Eq. (\ref{rhosep}) indicates that in order to describe non-separable states of indistinguishable fermions we need to resort to more general basis $\{\ket{\psi^{-}_{k}}_{S}\}$ that include elements different from Slater determinants.

Since there are $n(n-1)/2$ different $k'$s and $n=2s+1$, with $s$ being the spin of the particle, the dimension of $\mathcal{H}_{-}$ equals $s(2s+1)$. For $s=1/2$ the basis $\{\ket{\psi^{-}_{k}}_{S}\}$ possess a single element, it thus possess Slater rank 1, and hence no entanglement is present. Therefore the fermion system of lowest dimensionality exhibiting the phenomenon of entanglement corresponds to $s\geqslant3/2$, for which $n\geqslant4$ and dim$\mathcal{H}_{-}\geqslant6$. Denoting with $\{\ket{i}\}=\{\ket{1}, \ket{2}, ..., \ket{n}\}$ an orthonormal basis of the $n$-dimensional Hilbert space of each subsystem, we can identify each $\ket{i}$ with the states $|s, m_s \rangle $, with $m_s=-s \ldots, s,$ \cite{ESBL02} so that 
\beq
\{\ket{1}=\ket{s,s}, \ket{2}=\ket{s,s-1}, ..., \ket{n}=\ket{s,-s}\}. 
\eeq
Within this angular momentum representation, the antisymmetric joint eigenstates $\{ |j, m\rangle, \,\, -j\le m
\le j, \,\, 0\le j\le 2s \}$ of the total angular momentum
operators $J_z$ and $J^2$ constitute a natural basis $\{\ket{\psi^{-}_{k}}_{S}\}$ for the
Hilbert space associated with the pair of identical fermions. Such antisymmetric states are those characterized
by an even value of the quantum number $j$ \cite{F62,D02}. In what follows the 
notation $|j,m\rangle$ is always meant to refer to the angular 
momentum representation.\\

\noindent
The following is a list of the antisymmetric total angular 
momentum eigenstates for two fermions of spin $\frac{3}{2}$ with the value for the 
concurrence (see equation (\ref{concurrence})) indicated on the right (we use a compact notation according to which, for instance, the ket $|0, 0\rangle$ stands for $|j=0, m=0\rangle$):

\begin{center}
\begin{tabular}{l|c}
&$C$\\ \hline
$\ket{\psi^{-}_{1}}=|2,2\rangle = \frac{1}{\sqrt{2}}|12\rangle-\frac{1}{\sqrt{2}}|21\rangle $ & 0\\
$\ket{\psi^{-}_{2}}=|2,1\rangle =  \frac{1}{\sqrt{2}}|13\rangle-\frac{1}{\sqrt{2}}|31\rangle $ & 0\\
$\ket{\psi^{-}_{3}}=|2,0\rangle = \frac{1}{2}|23\rangle + \frac{1}{2}|14\rangle-\frac{1}{2}|41\rangle - \frac{1}{2}|32\rangle$ & 1 \\
$\ket{\psi^{-}_{4}}=|2,- 1\rangle = \frac{1}{\sqrt{2}}|24\rangle-\frac{1}{\sqrt{2}}|42\rangle $ & 0\\
$\ket{\psi^{-}_{5}}=|2,- 2\rangle = \frac{1}{\sqrt{2}}|34\rangle-\frac{1}{\sqrt{2}}|43\rangle $ & 0\\
$\ket{\psi^{-}_{6}}=|0,0\rangle = \frac{1}{2}|32\rangle - \frac{1}{2}|23\rangle + \frac{1}{2}|14\rangle-\frac{1}{2}|41\rangle $ & 1 \\
\end{tabular}
\end{center}
Notice that the states $|0,0\rangle$ and $|2,0\rangle$ are maximally entangled, 
while all the other states in the list correspond to single Slater determinants thus 
have zero entanglement.\\

\noindent
Necessary and sufficient separability criteria for pure states of two identical fermions have been formulated in
terms of appropriate entropic measures evaluated on the
single-particle reduced density matrix (see \cite{PMD09} and references therein). For mixed states, however, the development of entanglement criteria and
entanglement measures remains largely unexplored. Only for fermionic systems described by a single-particle
Hilbert space of dimension 4 a closed
analytical expression for the amount of entanglement, or concurrence $C(\rho_{ab})$, in a general (pure or mixed) two-fermion state $\rho_{ab}$ is
known \cite{ESBL02},

\begin{equation}\label{concurrence}
 C(\rho_{ab})=\textrm{max}\{0,\lambda_1-\lambda_2-\lambda_3-\lambda_4-\lambda_5-\lambda_6\},
\end{equation}

\noindent
where the $\lambda_i$'s are, in decreasing order, the square roots of the eigenvalues
of $\rho_{ab}\tilde{\rho}_{ab}$ with $\tilde{\rho}_{ab}={\mathbb D}\rho_{ab} {\mathbb D}^{-1}$, where
${\mathbb D}$ is given by

\begin{equation}\label{matrixD}
{\mathbb D}=\left(
      \begin{array}{cccccccc}
            0 & 0 & 0 & 0 & 1 & 0 \\
            0 & 0 & 0 & -1 & 0 & 0 \\
        0 & 0 & 1 & 0 & 0 & 0 \\
        0 & -1 & 0 & 0 & 0 & 0 \\
            1 & 0 & 0 & 0 & 0 & 0 \\
        0 & 0 & 0 & 0 & 0 & 1 \\
        \end{array}
      \right)\kappa,
\end{equation}

\noindent
$\kappa$ is the complex conjugation operator and ${\mathbb D}$ is expressed
with respect to the total angular momentum basis in the following order
$|2,2\rangle$, $|2,1\rangle$, $|2,0\rangle$, $|2,-1\rangle$, $|2,-2\rangle$,
and $i|0,0\rangle$.

\subsection{Entanglement between one fermion and the rest of the system}

Considering that the tripartite system consisting of two identical fermions and the environment $E$ is in a pure state $|\psi\rangle$, we now look for a quantitative indicator of the amount of entanglement between one of the fermions and the rest of the system. In order to do so we first notice the following:

\begin{itemize}
 \item Since both fermions are identical, the amount of the entanglement exhibited by one of them with the rest of the system must be the same for both of them.
 \item If each fermion can be regarded as disentangled from the rest, then the fermions pair as a whole, is disentangled from the environment $E$.
\end{itemize}

It is natural to use as an indicator of the amount of entanglement of one fermion with the rest an entropic measure evaluated on the single-fermion reduced density matrix $\rho_{f}$ (here $f$ is either $a$ or $b$), obtained after tracing the full state $\rho=|\psi\rangle\langle\psi|$ over one fermion and over the environment. Let us consider the von Neumann entropy 

\begin{equation}
 S_{vN}(\rho_f)=-\textrm{Tr}(\rho_f \ln\rho_f).
 \end{equation}

Now, the global state $|\psi\rangle$ can be expressed in terms of the Schmidt decomposition

\begin{equation}
 |\psi\rangle=\sum_i\sqrt{\lambda_i}|\phi^-_{i}\rangle|E_i\rangle,
\end{equation}
where $\lambda_i$ are the Schmidt coefficients such that $\sum_i\lambda_i=1$, and $\{|\phi^-_{i}\rangle\}$ and $\{|E_i\rangle\}$ are sets of orthonormal states belonging, respectively, to $\mathcal{H}_-$ and $\mathcal{H}_E$. Let $\rho_{fi}$ denote the single-particle density matrix corresponding to the two-fermion state $|\phi^-_{i}\rangle$. Then,

\begin{equation}
 \rho_f=\sum_i\lambda_i\rho_{fi}.
\end{equation}
Using the concavity of $S_{vN}$ we get

\begin{equation}\label{concSvN}
 S_{vN}(\rho_f)\geq\sum_i\lambda_i S_{vN}(\rho_{fi}).
\end{equation}

On the other hand, it holds that \cite{PMD09}
\begin{equation}\label{maxSvNi}
 S_{vN}(\rho_{fi})\geq\ln 2.
\end{equation}
Therefore, combining (\ref{concSvN}) and (\ref{maxSvNi}) we obtain,
\begin{equation}\label{maxSvN}
  S_{vN}(\rho_f)\geq\sum_i\lambda_i S_{vN}(\rho_{fi})\geq\ln 2.
\end{equation}
The equality sign in (\ref{maxSvNi}) holds only if each $|\phi^-_{i}\rangle$ corresponds to one-single Slater determinant. The equality in (\ref{concSvN}) occurs only if all the $\rho_{fi}$ are equal to each other \cite{Wehrl78}. Combining these two conditions it follows that the lower bound in (\ref{maxSvN}), $S_{vN}(\rho_f)=\ln2$, happens only if the two fermions are disentangled from the environment and disentagled from each other. That is, when the two-fermion state is described by a Slater determinant.

The above considerations show that the lower bound in (\ref{maxSvN}) corresponds to the physical situation in which each of the fermions has to be regarded as disentangled form the rest of the system. Also, the quantity
\begin{equation}
 \varepsilon=S_{vN}(\rho_f)-\ln 2,
\end{equation}
provides a useful quantitative indicator of the amount of entanglement between one fermion and the rest of the system. Notice that $\varepsilon$ vanishes if and only if the two fermions are disentangled from the environment and disentagled from each other. The measure $\varepsilon$ has a non-zero value if the two fermions are entangled with each other, or entangled with the environment, or both.

The measure $\varepsilon$ is different from a measure of entanglement between the two fermions, and is also different from a measure of the entanglement between two fermions (as a whole) and the environment. In addition, $\varepsilon$ is fully consistent with previous approaches to entanglement between identical fermions (see, for instance \cite{GMW02}). In particular, it takes into account the fact that the minimum correlations required by anti-symmetry do not contribute to entanglement.

Basically the same reasoning used above can be applied to argue that
\begin{equation}
 \frac{1}{2}-\textrm{Tr}(\rho_f^2)
\end{equation}
is an appropriate quantifier of the entanglement between one fermion and the rest of the system. This is based on the fact that $1-\textrm{Tr}(\rho_f^2)$ is a concave functional of $\rho_f$, and that for a pure state $|\phi\rangle$ of two fermions we always have $1-\textrm{Tr}(\rho_f^2)\geq\frac{1}{2}$ with equality if and only if $|\phi\rangle$ is a Slater determinant \cite{PMD09}. Consequently, we can resort to
\begin{equation}\label{Cf}
C_{a|Eb}=\sqrt{\frac{2d}{d-2}\left(\frac{1}{2}-\textrm{Tr}\rho^{2}_{a}\right)}
\end{equation}
to quantify the entanglement between one of the fermions and the rest, when the two-fermion plus environment system is in a pure state. Here $d\geq 4$ is the dimension of the single-fermion Hilbert space, and the factor $2d/(d-2)$ is introduced so that $C_{a|Eb}$ lies between 0 and 1.

\subsection{Entanglement between one fermion and the environment}\label{aE}

For the tripartite system $abE$, let us consider the observables of the form,

\begin{equation} \label{1e_observable1}
{\cal O} = \frac{1}{2} \sum_i \Bigl(A_i \otimes \mathbb{I} +
\mathbb{I}\otimes A_i \Bigr) \otimes B_i.
\end{equation}

\noindent
In this equation the $A_i$'s act on the single-fermion Hilbert space, that is,
they correspond to observables representing properties of
one single fermion.  $\mathbb{I}$ is the identity operator acting on the single-fermion
space, and the $B_i$'s are observables referred to the environment.
The expectation value of ${\cal O}$ in the tripartite state $\rho$ reads

\begin{eqnarray}    \label{1e_observable2}
\langle {\cal O} \rangle &=& \textrm{Tr}\left\{\left[\frac{1}{2} \sum_i \Bigl(A_i \otimes \mathbb{I} +
\mathbb{I}\otimes A_i \Bigr) \otimes B_i \right] \rho \right\}  \cr &=&
\textrm{Tr} \left[\left( \sum_i A_i \otimes B_i \right) \rho_{fE} \right],
\end{eqnarray}

\noindent
where $\rho_{fE}$ is the density matrix obtained after tracing the global density matrix
over the degrees of freedom of one fermion, e.g., $\rho_{aE} = \textrm{Tr}_b\rho.$

The observables of the form (\ref{1e_observable1}) are those representing properties referred to
one fermion and the environment. Equation (\ref{1e_observable2}) means that, as far as these
observables are concerned, all the statistics associated with quantum measurements are
described by the reduced density matrix $\rho_{fE}$. That is, $\rho_{fE}$ jointly describes single-fermion and environment features of the system,
 including the concomitant correlations (both quantum and classical) between single-fermion
 properties and environment properties. Consequently, it is physically meaningful to regard
 the entanglement of the state $\rho_{fE}$ (measured in the usual sense when considering
 distinguishable subsystems) as describing the entanglement between one fermion and the environment.
 Therefore, the entanglement between one fermion and the environment can be operationally defined as the
 effective entanglement between the fermions and the environment when only single-fermion
 properties can be measured.
 
As we are dealing with pure global (tripartite) states $\rho$, the reduced density matrix $\rho_{fE}$ will in general be a mixed state of an $n$-level and an $m$-level system, $m$ being the dimension of $\mathcal{H}_E$, whence we will use the negativity \cite{ZHSLK98,VW02} as an indicator of entanglement between one fermion and the environment. The negativity $\mathcal{N}$ is given by the sum of the negative eigenvalues $\alpha_i$ of the partial transpose (with respect to either $E$ or $f$) of the matrix $\rho_{fE}$,

\begin{equation}\label{Negativity}
 \mathcal{N}=\sum_i |\alpha_i|.
\end{equation}
By virtue of the PPT criterium, a positive value of $\mathcal{N}$ indicates that the state $\rho_{fE}$ is entangled.

\section{Decoherence process in two-fermion systems}

In this Section we shall analyse the dynamics of entanglement in an open system consisting of two identical fermions. According to the discussion in Section \ref{sim}, we focus on fermions that share a common environment, and that evolve under different decoherence processes when the initial state has the form (\ref{rho00}). We will compute analytically the entanglement between different parts of the complete (fermion+environment) system, restricting the study to fermions with a single-particle Hilbert space of dimension four, immersed in a two-level environment $E$ whose states are $\ket{0},\ket{1}$. 

In particular, we will resort to Eqs. (\ref{concurrence}) and (\ref{Cf}) to compute the entanglement between the fermions ($C_{ab}$), and the entanglement between one fermion and the rest of the system ($C_{a|Eb}$), respectively. In addition, we will use the expression
\begin{equation}\label{Cijk}
C_{E|ab}=\sqrt{2(1-\textrm{Tr}\rho^{2}_{E})}=\sqrt{2(1-\textrm{Tr}\rho^{2}_{ab})},
\end{equation}
to calculate the entanglement between $E$ and the fermionic subsystem. Recall that equation (\ref{Cijk}) provides indeed an entanglement measure whenever the total state is pure, i.e., described by a vector $|\psi\rangle_{abE}$ \cite{PRA64}. As for the entanglement between one fermion and the environment, we will proceed as explained in Section \ref{aE} and use the negativity (\ref{Negativity}) to detect the entanglement between $f=a,b$ and $E$.

In the study of decoherence processes, the Kraus representation introduced in Section \ref{sim} is particularly useful, since it allows to represent the unitary evolution of the fermion ($S$) plus environment ($E$) system by the quantum map \cite{Luiz}
\begin{equation}
\ket{\phi^{-}_{k}}_{S}\ket{0}_{E}\rightarrow (K_{0}\ket{\phi^{-}_{k}}_{S})\ket{0}_{E}+(K_{1}\ket{\phi^{-}_{k}}_{S})\ket{1}_{E}.
\end{equation}
Two paradigmatic quantum channels, widely used in studying the decoherence in open qubit systems, are the Amplitude Damping Channel (ADC) and the Phase Damping Channel (PDC). The former represents the dissipative interaction between the qubit and its environment, and the later can represent the coupling of the system to a noisy environment \cite{NC00}. Here we will generalize the main features of these channels to extend the corresponding quantum map to the 6-dimensional joint Hilbert space of the two fermions.
 
\subsection{Amplitud Damping Channel}

\begin{figure}
\begin{center}
\includegraphics[scale=0.3,angle=270]{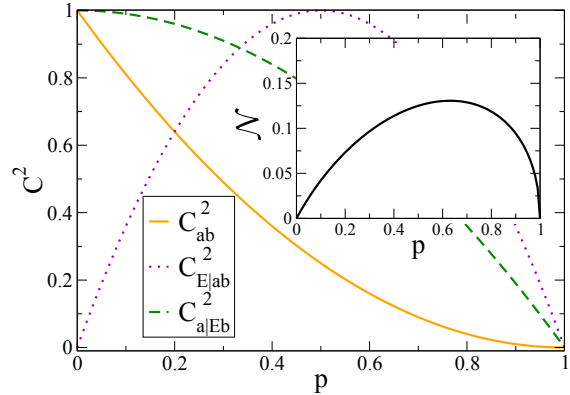}
\caption{(Color online). Entanglement evolution for the initial state (\ref{phi0AD}), under Amplitud Damping decoherence channel. Evolution of the entanglement between the fermions (orange solid line), evolution of the entanglement between the environment and the pair of fermions (violet dotted line), and evolution of the entanglement between one fermion and the rest of the system (green dashed line). Inset: evolution of the negativity of the reduced state of one fermion and the environment. All depicted quantities are dimensionless. \label{figure1}}
\end{center}
\end{figure}

The main feature of the ADC is that it preserves the total (system plus environment) excitation number. Considering that in the bipartite states $|2,m\rangle$, $m$ stands for an excitation that can be exchanged with the environment, the AD map in this case reads 
\begin{eqnarray}\label{ADchannel}
|2,m\rangle_{S}|0\rangle_E&\rightarrow&\sqrt{1-p}|2,m\rangle_{S}|0\rangle_E+\sqrt{p}|2,m-1\rangle_{S}|1\rangle_E\nonumber\\&&m=-1,...,2,\nonumber\\
|0,0\rangle_{S}|0\rangle_E&\rightarrow&|0,0\rangle_{S}|0\rangle_E,\nonumber\\
|2,-2\rangle_{S}|0\rangle|_E&\rightarrow&|2,-2\rangle_{S}|0\rangle_E,
\end{eqnarray}
where $p\in [0,1]$ is a continuous parameter characterizing the evolution. Let us consider the initial state
\begin{equation}
 |\psi(0)\rangle_{SE}=|2,0\rangle_{S}|0\rangle_E.
 \label{phi0AD}
\end{equation}
According to Eq. (\ref{ADchannel}), the whole tripartite system evolves to
\begin{equation}
 |\psi(p)\rangle_{SE}=\sqrt{1-p}|2,0\rangle_{S}|0\rangle_E+\sqrt{p}|2,-1\rangle_{S}|1\rangle_E. \label{30}
\end{equation}
We obtain the following expressions for the squared concurrences (tangles) as a function of $p$:
\begin{eqnarray}
 C^2_{ab}(p)&=&(1-p)^2\label{Cab},\\
 C^2_{E|ab}(p)&=&4p(1-p)\\
 C^2_{a|Eb}(p)&=&1-p^2,
\end{eqnarray}
and plot them in Fig. \ref{figure1}. The solid (orange) line shows a typical feature of decoherence processes: as a result of the interaction of the fermionic system with the environment, $C^2_{ab}$ decreases monotonically until its completely disappearance. The dotted (purple) curve, representing $C^2_{E|ab}$, shows that along the evolution the environment gets entangled with the fermionic system, disentangling from it only at $p=1$. It is also observed that the bipartite entanglement between $a$ and the rest ($Eb$) (green dashed curve) decreases at a slower rate than $C^2_{ab}$, with $C^2_{a|Eb}\geqslant C^2_{ab}$.
In the inset of Fig. \ref{figure1} we plot the evolution of the negativity $\mathcal{N}(\rho_{fE})$. Such quantity is positive for all $p\in(0,1)$, indicating that in this interval there exist entanglement between each single fermion and the environment.

The dynamics of entanglement induced by the ADC has been previously studied in the context of two initially entangled (distinguishable) qubits, $q_1$ and $q_2$, when only $q_2$ interacts (locally) with its environment $E_2$ (see, e.g., Ref. \cite{AguilarPRA14}). Though in the present fermionic system the environment is common to both particles so the evolution is non-local, the comparison between the distinguishable-qubit and the identical-fermion case seems useful to evince the main features that distinguishes one and the other type of evolution. In particular, in the qubit case, it is found \cite{AguilarPRA14} that as a result of the decoherence channel, there is an entanglement swapping between $q_{1}q_{2}$ and $q_{1}E_2$, that is, the initial ($p=0$) entanglement  between $q_{1}$ and $q_{2}$ is completely converted (at $p=1$) into entanglement between $q_{1}$ and $E_{2}$. In this sense the net effect of the ADC is to redistribute and transfer the initial bipartite entanglement without looses. In the fermion case this no longer holds. This can be seen by taking $p=1$ in the state (\ref{30}) and observing that since $|2,-1\rangle$ is a two-fermion separable state (see the Table), the final tripartite state $|\psi(1)\rangle_{SE}=|2,-1\rangle_{S}|1\rangle_E$ is completely disentangled. Thus the initial (maximal) entanglement between the fermions is finally lost due to the decoherence process, yet during the evolution (i.e., for $0<p<1$) the entanglement redistribution due to the ADC is of course present, as seen in Fig. \ref{figure1}. These observations would thus indicate that the open qubit system is more robust against decoherence than the identical-fermion system. This is reinforced by the fact that for an initial maximally entangled state in the qubit case we have \cite{AguilarPRA14} $C^2_{q_{1}q_2}(p)=1-p$, whereas here the tangle between the two fermions is given by Eq. (\ref{Cab}), namely $C^2_{ab}(p)=(1-p)^2$, so the entanglement between the fermions decays faster than the entanglement between the qubits.
  
As a second example we consider the initial state:
\begin{equation}\label{phi1AD}
 |\phi(0)\rangle_{SE}=(\alpha|2,1\rangle_{S}+\beta|2,-1\rangle_{S})|0\rangle_E,
\end{equation}
with $|\alpha|^2+|\beta|^2=1$. Applying the map (\ref{ADchannel}) the state evolves to
\begin{eqnarray}
|\phi(p)\rangle_{SE}&=&\alpha[\sqrt{1-p}|2,1\rangle_{S}|0\rangle_E+\sqrt{p}|2,0\rangle_{S}|0\rangle_E]\\&+&\beta[\sqrt{1-p}|2,-1\rangle_{S}|0\rangle_E+\sqrt{p}|2,-2\rangle_{S}|1\rangle_E],\nonumber
\end{eqnarray}
and in this case we obtain:
\begin{eqnarray}\label{ab}
 C^2_{ab}(p)&=&\left[(1-p)(\alpha\beta^*+\alpha^*\beta)-p|\alpha|^2\right]^2\nonumber,\\
 C^2_{E|ab}(p)&=&4p(1-p)\nonumber,\\
 C^2_{a|Eb}(p)&=&2-4p|\alpha|^2|\beta|^2-[1+(1-p)^2]|\alpha|^4\nonumber\\&-&2[p+(1-p)^2]|\beta|^4.
\end{eqnarray}
Three qualitatively different cases will be now analysed for different values of the parameters $\alpha$ and $\beta$. For $\alpha=0$ ($\beta=1$) we get $C^2_{ab}(p)=0$, thus the fermions remain in a separable state along the whole evolution, whose only effect is that of continuously transforming the state $\ket{2,-1}$ (at $p=0$) into $\ket{2,-2}$ (at $p=1$) without modifying the entanglement between the identical parties. 
\begin{figure}
\begin{center}
\includegraphics[scale=0.3,angle=270]{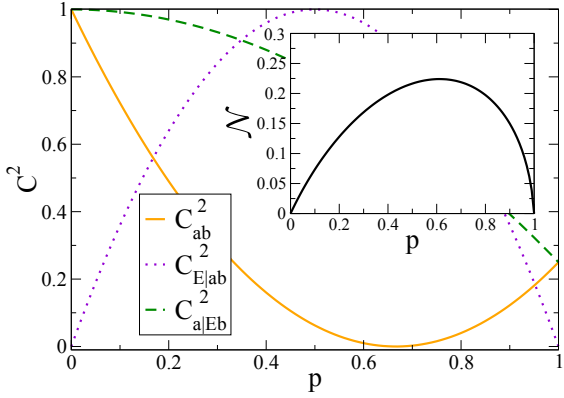}
\caption{(Color online). Entanglement evolution for the initial state (\ref{phi1AD}) with $\alpha=\beta=1/{\sqrt{2}}$ under Amplitud Damping decoherence channel. Evolution of the entanglement between the fermions (orange solid line), evolution of the entanglement between the environment and the two fermions (violet dotted line), and evolution of the entanglement between one fermion and the rest of the system (green dashed line). Inset: evolution of the negativity of the reduced state of one fermion and the environment. All depicted quantities are dimensionless.\label{figure2}}
\end{center}
\end{figure}
For $\alpha=\beta=1/{\sqrt{2}}$, we obtain the following:
\begin{eqnarray}
C^2_{ab}(p)&=& \left(1-\frac{3}{2}p\right)^2,\nonumber\\
C^2_{E|ab}(p)&=& 4p(1-p)\nonumber,\\
C^2_{a|Eb}(p)&=&1-\frac{3}{4}p^2.
\end{eqnarray}
Thus $C^2_{ab}(p)$ decreases monotonically from its maximum value to zero, at $p=2/3$, where it starts to increase as shown in Fig. \ref{figure2}. Such increase in the entanglement between noninteracting entities is a result of the nonlocal dynamics due to the common (or global) environment \cite{Salles08,Luiz}. A more drastic example of the increase in the entanglement between the fermions due to the global environment can be seen by taking $\beta=0$ ($\alpha=1$) in Eqs. (\ref{ab}). In this case the initial state $\ket{2,1}$ is separable (see Table), but as $p$ increases the entanglement between fermions increases as well, since $C^2_{ab}(p)=p^2$, so that at $p=1$ the fermions end up maximally entangled. In other words, when applied to appropriate initial states, the decoherence channel is capable of increasing the entanglement between the fermions. 

The negativity for the case $\alpha=\beta=1/{\sqrt{2}}$, shown in the inset of Fig. \ref{figure2}, is qualitatively the same as in the previous case (initial state (\ref{phi0AD}) subject to the ADC). Again, in the interval $p\in(0,1)$ there exist a nonzero entanglement between one fermion and the environment.


\subsection{Phase Damping Channel}

This process describes the loss of quantum information with probability $p$ without any exchange of energy. The PDC is described by the quantum map

\begin{eqnarray}
|j,m\rangle_{S}|0\rangle_E&\rightarrow&\sqrt{1-p}|j,m\rangle_{S}|0\rangle_E+\sqrt{p}|j,m\rangle_{S}|1\rangle_E\nonumber\\&&j=2,m=-2,...,2\nonumber\\
|0,0\rangle_{S}|0\rangle_E&\rightarrow&|0,0\rangle_{S}|0\rangle_E.
\end{eqnarray}

We apply the map to the initial state
\begin{equation}
 |\eta(0)\rangle_{SE}=(\delta|2,0\rangle_{S}+i\gamma|0,0\rangle_{S})|0\rangle_E,\label{DC1}
\end{equation}
with $|\delta|^2+|\gamma|^2=1$. The evolved state reads
\beq\label{eta(p)}
\ket{\eta(p)}=\delta\ket{2,0}_S\ket{P(p)}_E+i\gamma\ket{0,0}_S\ket{0}_E,
\eeq
where we defined
\beq
\ket{P(p)}_E=\sqrt{1-p}\ket{0}_E+\sqrt{p}\ket{1}_E.
\eeq

\begin{figure}[h]
\begin{center}
\includegraphics[scale=0.3,angle=0]{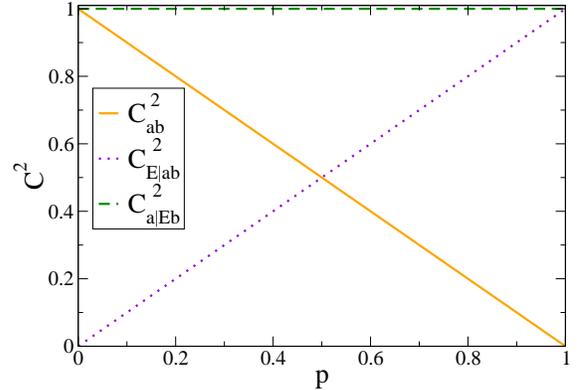}
\caption{Entanglement evolution for the initial state (\ref{DC1}) with  $\delta=1/\sqrt{2}, \gamma=-i\delta$ under Phase Damping decoherence channel. Evolution of the entanglement between the fermions (orange solid line), evolution of the entanglement between the environment and the two fermions (violet dotted line), and evolution of the entanglement between one fermion and the rest of the system (green dashed line). All depicted quantities are dimensionless.\label{figure3}}
\end{center}
\end{figure}

The squared concurrences for this case are given by
\begin{eqnarray}
 C^2_{ab}(p)&=&\zeta(p,\delta,\gamma)\nonumber\\&-&\sqrt{\zeta^2(p,\delta,\gamma)-[\zeta(p,\delta,\gamma)-2p|\delta|^2|\gamma|^2]},\nonumber\\
 C^2_{E|ab}(p)&=&2[1-\zeta(p,\delta,\gamma)]\nonumber,\\
 C^2_{a|Eb}(p)&=&1-(1-p)(\delta^*\gamma-\delta\gamma^*)^2,
\end{eqnarray}
where $\zeta(p,\delta,\gamma)=|\delta|^4+|\gamma|^4+2(1-p)|\delta|^2|\gamma|^2$. Setting $\delta=1/\sqrt{2}, \gamma=-i\delta$, we get the following expressions:
\begin{eqnarray}\label{concuDC}
 C^2_{ab}(p)&=&1-p,\nonumber\\
 C^2_{E|ab}(p)&=&p,\nonumber\\
 C^2_{a|Eb}(p)&=&1,
\end{eqnarray}
and plot them in Fig. \ref{figure3}. Now $C^2_{ab}$ decreases linearly in $p$ whereas $C^2_{E|ab}$ increases at the same rate, so the sum $C^2_{ab}+C^2_{E|ab}$ remains constant along the evolution, and equal to $C^2_{a|Eb}$. Unlike the previous (ADC) case, here the environment ends up being maximally entangled with the bipartite system $S$ at the expense of the disentangling between the fermions.

As for the entanglement between one fermion and the environment, we resort to Eq. (\ref{eta(p)}) to obtain the reduced density matrix $\rho_{aE}(p) = \textrm{Tr}_b\ket{\eta(p)}\bra{\eta(p)}$. Direct calculation leads to 
\begin{eqnarray}\label{sep}
 \rho_{aE}(p)&=&\frac{1}{4}[\ket{1}\bra{1}_a+\ket{4}\bra{4}_a]\ket{\varphi_+(p)}\bra{\varphi_+(p)}_E\nonumber\\&+&\frac{1}{4}[\ket{2}\bra{2}_a+\ket{3}\bra{3}_a]\ket{\varphi_-(p)}\bra{\varphi_-(p)}_E,
 \end{eqnarray}
where $\ket{i}_a$ ($i=1,2,3,4$) are the single-fermion states defined in Section \ref{IIIA}, and $\ket{\varphi_\pm(p)}$ stands for the vector
\begin{eqnarray}
\ket{\varphi_\pm(p)}&=&\frac{1}{\sqrt{2}}[\ket{0}\pm\ket{P(p)}]\nonumber\\&=&\frac{1}{\sqrt{2}}[(1\pm\sqrt{1-p})\ket{0}\pm\sqrt{p}\ket{1}].
\end{eqnarray}
Equation (\ref{sep}) shows that the state $\rho_{aE}$ is separable for all $p$, so that any measure $C_{aE}$ quantifying the entanglement between one fermion and the environment vanishes, i.e., $C_{aE}(p)=0$. 

It is interesting to observe that this latter result, together with Eqs. (\ref{concuDC}), coincide with the qubit concurrences obtained in the previously discussed 3-qubit system \cite{AguilarPRA14}. Thus, contrary to what happened in the ADC case, the effect of the PDC on both (qubit and fermion) systems seems to be the same regardless of the local (or non-local) nature of the interaction. In order to go further in the comparison between the distinguishable-qubit and the identical-fermion case, we recall the monogamy inequality

\begin{equation}\label{monoq}
C_{i|jk}^ 2-C_{ij}^2-C_{ik}^2\geqslant 0
\end{equation}
satisfied by the usual concurrence, i.e., involving distinguishable qubits $i,j$ and $k$ \cite{CKW00}. Motivated by this inequality we define
\begin{equation}\label{Ra}
R_a=C_{a|Eb}^2-C_{aE}^2-C_{ab}^2
\end{equation}
and
\begin{equation}\label{RE}
 R_E=C_{E|ab}^ 2-C_{Ea}^ 2-C_{Eb}^2,
\end{equation}
where $C_{aE}=C_{Ea}=C_{Eb}$ is an appropriate measure (consistent with the previously defined concurrences) of the entanglement between the two-level environment and the four-level fermion. Using $C_{fE}(p)=0$ and Eqs. (\ref{concuDC}) leads to
\beq\label{ResDC}
 R_{a}(p)=R_{E}(p)=r(p)=p\geqslant0.
 \eeq

Since $R_{a,E}$ encodes information of the entanglement that cannot be written as entanglement between two parties (hence reflect multipartite entanglement), a positive value of $R_{a,E}$ exhibits the presence of tripartite entanglement. Moreover, since $R_{a}=R_{E}$, such tripartite entanglement is the same in all bipartitions (fermion$|$rest, environment$|$rest). In the 3-(distinguishable) qubit system the corresponding residual entanglement is just the 3-tangle $\tau_{ijk}=\tau=C_{i|jk}^ 2-C_{ij}^2-C_{ik}^2$, which measures the genuine tripartite entanglement of those states pertaining to the GHZ-type family \cite{DVC00}. Thus, we can say that $r(p)$ here measures the genuine tripartite entanglement shared by the two fermions and the environment, and that the state (\ref{eta(p)}) is the analogous of the GHZ-family states for systems involving two fermions and their common environment. According to Eq. (\ref{ResDC}), the genuine tripartite entanglement increases linearly in $p$, and becomes maximum at $p=1$, where the `fermion-environment GHZ' state,
\beq\label{ghz}
\ket{GHZ}_{SE}=\ket{\eta(1)}=\frac{1}{\sqrt{2}}(\ket{2,0}_S\ket{1}_E+\ket{0,0}_S\ket{0}_E),
\eeq
 is reached. The state (\ref{ghz}) shares with the usual (3-qubit) GHZ state the property of having maximal genuine entanglement, while having zero entanglement between the parties when one of them (any) is traced out.
  
\section{Entropic entanglement criteria}

As we mentioned before, the particular case of systems of two identical fermions with a four-dimensional single-particle
Hilbert space (the simplest fermion system admitting entanglement) is the only one for which we have a closed, analytical expression for the concurrence. No such expression is known for fermion systems of higher dimensionality.

In order to study the entanglement dynamics of systems of $N$-fermions undergoing decoherence it is possible to use an entanglement indicator based upon entropic criteria \cite{ZPCP12}. In this section we will use our previous results for the case of systems of dimension 4 to investigate the efficiency of these criteria.

All separable states (pure or mixed) of $N$ identical fermions comply with the entropic inequalities
\begin{equation}
 S_{R}^{(\alpha)}(\rho_F)+\ln N\geq S_{R}^{(\alpha)}(\rho_f),
\end{equation}
where $S_{R}^{(\alpha)}$ is the Rényi entropy with $\alpha\geq 1$, $\rho_{F}$ is the global $N$-fermions density matrix, and $\rho_f$ is the single-particle reduced density matrix. The equality sign in the above inequality occurs, for instance, in the case of pure separable states. Now, if ones considers the quantity 
\begin{equation}
 Q^{(\alpha)}(\rho_F)=S_{R}^{(\alpha)}(\rho_f)-S_{R}^{(\alpha)}(\rho_F)-\ln N,
\end{equation}
then for all separable states (pure or mixed) one has 
\begin{equation}
 Q^{(\alpha)}(\rho_F)\leq 0.
\end{equation}
Therefore for $Q^{(\alpha)}>0$ one knows for sure that the state is entangled:
\begin{equation}
 Q^{(\alpha)}(\rho_F)>0\Rightarrow \rho_F\; \text{entangled}. 
\end{equation}
Since the converse ($\rho_F\; \text{entangled}\Rightarrow Q^{(\alpha)}>0$) does not hold in general,  the condition $Q^{(\alpha)}>0$ detects some (mixed) entangled states but not all of them. 
For two-fermion states the entanglement criterion improves as $\alpha$ increases and is the most efficient in the limit $\alpha\to\infty$ \cite{ZPCP12}. Note that the criterion associated with the von Neumann entropy constitutes a special instance, corresponding to the particular value $\alpha\to 1$ of the R\'enyi entropic parameter.
\begin{figure}[h]
\begin{center}
\includegraphics[scale=0.3,angle=270]{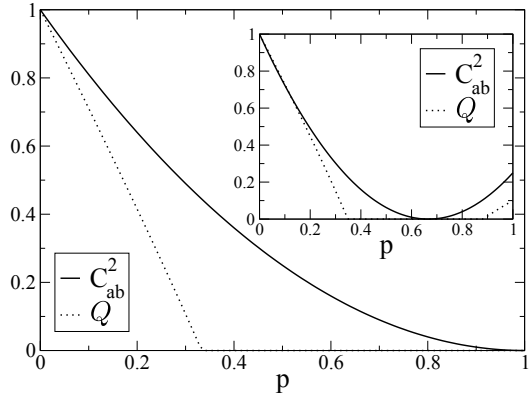}
\caption{Entropic entanglement indicator evolution (dashed line), and evolution of the fermionic concurrence (solid line) for the initial state (\ref{phi0AD}) under ADC. Inset: Evolution of the entropic entanglement indicator (dashed line) and fermionic concurrence dynamics (solid line) for the initial state (\ref{phi1AD}) with $\alpha=\beta=1/{\sqrt{2}}$ under ADC. The logarithms in the entropic entanglement indicator are taken to the base 2 and all depicted quantities are dimensionless.\label{figure4}}
\end{center}
\end{figure}

The quantity $Q^{(\alpha)}$ can thus be regarded as an entanglement indicator whose evolution under decoherence can be investigated. In particular, we study the evolution of $Q^{(\alpha)}$ under decoherence for two previously considered examples setting $\alpha=\infty$. In this case we have (with $Q^{(\infty)}=Q$)
\begin{equation}
 Q(\rho_F)=S_{R}^{(\infty)}(\rho_f)-S_{R}^{(\infty)}(\rho_F)-\ln 2,
\end{equation}
with
\begin{equation}
 S_{R}^{(\infty)}(\rho_F)=-\ln \lambda_{max},
\end{equation}
where $\lambda_{max}$ is the largest eigenvalue of $\rho_F$. Figure \ref{figure4} shows the evolution of $Q$ (dashed lines) for the initial state (\ref{phi0AD}) under the Amplitude Damping decoherence channel. It is observed that the evolution of the entanglement indicator is qualitatively the same as the evolution of the fermionic concurrence (solid lines). The resemblance is stronger in the case of the initial state (\ref{phi1AD}) with $\alpha=\beta=1/{\sqrt{2}}$ (inset). However, in any case, we can conclude that $Q$ is a reasonably good entanglement detector.

The entanglement indicators $Q^{(\alpha)}$, as well as the entanglement measures
considered in Section III, are not straightforwardly measurable, in the sense 
that they are not equal to (or function of a small number of) expectation values 
of quantum mechanical observables. However, if the global state of the two-fermion
system under consideration is first reconstructed via appropriate quantum state 
tomography techniques, then the aforementioned entanglement quantities 
can also be experimentally determined.  More directly measurable entanglement indicators
for fermionic systems have not yet been as intensively investigated  as those 
for systems with distinguishable parts. However, some progress in this direction 
has been done. For instance, the entanglement indicators advanced in \cite{ZP10},
based upon uncertainty relations, are expressed in terms of expectation values of 
measurable quantum observables.   

\section{Conclusions}

We studied the effects arising from the interaction between a quantum system of two identical fermions and the environment. We showed that for the exchange symmetry to be preserved, the evolution of the system must be global, or nonlocal, in the sense that each fermion interacts separately with a \textit{common} environment. Thus, in order to analyse the dynamics of entanglement under two paradigmatic decoherence channels widely studied in the context of local qubit dynamics, we generalised and extended the Amplitude Damping Channel and the Phase Damping Channel to the joint Hilbert space of the two fermions. 

In order to achieve a more complete analysis of the evolution of the entanglement in the tripartite system (fermion+fermion+environment), it was necessary to define two measures of entanglement: one that quantifies the entanglement between one fermion and the rest of the system (fermion+environment), and one that quantifies the entanglement between one (any) of the fermions and the environment. With these tools we were able to study the dynamics of entanglement for some initially entangled states subject to the ADC and the PDC. Comparison with the 3-qubit case was made, and new insights into the mechanism of entanglement evolution in open systems of identical particles were revealed. 

In the case of the Phase Damping Channel, and by resource of a monogamy relation, we were able to detect genuine tripartite entanglement and determined an analogous to the GHZ-state involving the two fermions and the environment. Further progress in relation with tripartite entanglement and monogamy relations in these kind of systems is, however, inherently constrained by the advance in the problem of quantifying entanglement in systems involving identical particles, and multipartite systems in general, a problem that remains far from closed. 

Finally we showed the dynamics of an entanglement indicator based on an entropic criteria which can be used to study decoherence in more general (higher dimensions) systems of identical fermions. For the four-dimensional case studied here, the entanglement indicator turned out to be a reasonably good indicator of entanglement between the pair of fermions.

A possible experimental scenario in which to consider the kind of processes
discussed in the present work could be provided by a system consisting of two
electrons in two laterally coupled quantum dots \cite{SLM01}. This system allows for 
the implementation of quantum information related processes such as quantum gates,
and can be described in terms of an effective four dimensional single-particle
Hilbert space (leading to a six-dimensional two-fermion Hilbert space). The relevant
single-particle Hilbert space is spanned by single-electron states that, in self-explanatory
notation, can be expressed as   $  \{  | A  \uparrow   \rangle,  | A  \downarrow \rangle, 
| B  \uparrow  \rangle,   | B  \downarrow \rangle \}$,
where $\{  | A \rangle,   | B \rangle  \}$    denote two orthogonal electronic spatial 
wave functions (orbitals) predominantly located around two particular 
locations in the double quantum dot.  
This two-electron double quantum dot system then constitutes a possible experimental realization (of technological significance) of the type of two-fermion systems considered in the present contribution.

\begin{acknowledgements}
A.P.M. and A.V.H. acknowledge the Brazilian agencies MEC/MCTI/CAPES/CNPq/FAPs for the financial support through the \textit{BJT Ci\^encia sem Fronteiras} Program.
\end{acknowledgements}

\bibliographystyle{apsrev}
\bibliography{Master_Bibtex}

\end{document}